# Approaches to Maintain Bi-connectivity for Resilience in Overlaid Multicasting

Ashutosh Singh and Yatindra Nath Singh
*Department of Electrical Engineering,*
*Indian Institute of Technology,*
*Kanpur-208016, Uttar Pradesh, India*
{singhash, ynsingh}@iitk.ac.in

## Abstract

Application layer multicast (ALM) also called Overlay Multicast, is an attractive alternative solution to most of the problems associated with IP multicast. In ALM, multicast-related functionalities are moved to end-hosts. Application layer multicast builds a peer-to-peer (P2P) overlay topology consisting of end-to-end unicast connections between end-hosts. The key advantages, overlays offer, are flexibility, adaptability and ease of deployment [1]. The general approach to build an application layer multicast architecture involves tracking network characteristics and building appropriate topologies by allowing the end users to self organize into logical overlay networks for efficient data delivery. The major concern in designing ALM protocol is the mechanism to build and maintain a topology and to route data efficiently and reliably in this topology. We propose here a two-fold dynamic overlay tree construction and maintenance scheme in which a mesh-like topology is first built, where an arriving host connects to two already connected hosts. This ensures that two node and link disjoint paths are always maintained between every possible pair of nodes. Once the mesh is formed, on top of it, a single or multiple data delivery tree(s) are built using a suitable protocol. An algorithm is run in the nodes of the overlay topology to maintain the biconnectivity by inserting new links and deleting the redundant links as a continuous process.

**Key words:** Overlay network, Reliability, node disjoint paths, Biconnectivity

## I INTRODUCTION

The internet has seen an unprecedented growth due to the success of one-to-one applications such as reliable file transfer, electronic mail and World Wide Web (www). The Internet's unicast-only infrastructure does not provide efficient support for multicast applications viz. Internet-TV, multiparty Video conferencing, Live Lecture Delivery Systems (LLDS) and software distribution over Internet where multiple copies of a message need to be transported to multiple recipients at the different locations. IP multicast [2] at the network layer defines an efficient way for distribution of same content to multiple nodes. The sources transmit only one copy of the data and the appropriate network nodes make duplicate copies to forward on the distribution tree. However IP multicast has not been widely deployed [3]. IP multicast requires routers to maintain per-group state which leads to serious scaling constraints. IP multicast calls for changes at infrastructure level (need of multicast enabled routers) in the Internet which was originally designed for unicast applications. Further IP multicast is based on best-effort data delivery, and hence cannot support QoS. This is not desirable in applications such as multi-party gaming and multi-party video conferencing. Finally IP multicast does not provide solution for group management, multicast address allocation and support for network management. Since IP multicast is not a mandatory service in IPv4, most of the networks have not enabled multicast or provide it only as a value added service on a limited scale. So far, the multicast has been limited to islands of network domains under single administrative control or in the local area networks.

Application layer multicast (ALM) also known as Overlay Multicast is an attractive alternative solution, where multicast-related functionalities are moved to end-hosts. The key advantages, overlays offer, are flexibility, adaptability and ease of deployment. Application layer multicast builds a peer-to-peer (P2P) overlay topology consisting of end-to-end unicast connections between end-hosts. Multicast topology is maintained in this overlaid network. End users can constantly self organize the logical overlay networks for efficient data delivery. The shifting of multicast support from routers to end systems has the potential to address most problems associated with IP multicast. Since the multicast connections are based on end hosts, there is no need of multicast enabled routers.

However ALM incurs a performance penalty over IP Multicast. Links near the end users carry redundant copy of transmitted data. Also the delay to some of the end users is more than in case of IP multicast. The major concern in ALM is how to route data along the topology efficiently and reliably. To enhance the reliability, we propose creation and maintenance of a dynamic overlay tree built over bi-connected mesh overlay topology. In order to create and maintain bi-connected mesh, a new host connects to two already connected hosts in the overlay network. This results in at least two node-disjoint paths between any pair of nodes.

Once the mesh is formed, on top of that a data delivery tree is built using an appropriate distributed algorithm. Our simple broadcast based algorithm ensures the constant tracking of an alternate path (every node has knowledge of an alternate route to source) for data, so that if feed stops from the current path, it can switch over to the alternate path by sending the feed request in that direction. The algorithm ensures that the next best path to source is again identified. The detail of the algorithm is given in section 3.2.

A dynamic algorithm is run in the mesh to maintain biconnectivity by suitably adding new links when needed while also deleting the redundant links. The next section gives a brief survey of related work. Section III describes proposed approaches toward maintaining the biconnectivity. Finally we conclude in the section IV.

## II RELATED WORK

In this section, we describe the reliability schemes proposed in the major ALM protocols. In NARADA protocol, Chu *et. al.* [4], proposed a mesh based topology design for small group size. Data delivery trees are constructed entirely from the overlay links present in the mesh. Shortest path spanning trees are constructed per source by running distance vector algorithms on top of the mesh. Periodically, each member generates a refresh message with monotonically increasing sequence number and exchanges its knowledge of group membership with its neighbors in the mesh. Mesh partition is detected, when members on one side of the partition stop receiving sequence number updates from members on the other side. Each of such members is probed to determine if it is dead, else a link is added to it. Each of the members on one side may attempt to add new links to some partitioned member on other side; this situation is probabilistically resolved such that in spite of several members simultaneously attempting to repair partition, only a small number of new links are added.

To improve the mesh quality, dynamic addition and dropping of links is also done. Each member periodically probes some random member that is not a neighbor, and evaluates the utility of adding a link to this member. Link is added if expected utility gain exceeds a given threshold. To drop a link, members periodically compute the consensus cost of its link to every neighbor. The consensus cost of a link (i, j) is defined as max($cost_{ij}$, $cost_{ji}$), where $cost_{ij}$ is the number of members for which i uses j as next hop for forwarding packets, and $cost_{ji}$ is the number of members for which j uses i as next hop for forwarding packets. The link with lowest consensus cost is dropped if the consensus cost falls below a certain threshold.

In PRM enhanced NICE protocol, Banerjee *et. al.* [5] introduced multicast data recovery scheme called Probabilistic Resilient Multicast (PRM) with two components. A proactive component called randomized forwarding in which each overlay node chooses a constant number of other overlay nodes uniformly at random and forward data to each of them with a low probability and a reactive component called triggered NAKs to handle data losses due to link errors and network congestion.

## III OUR APPROACHES TOWARD BICONNECTIVITY FOR RESILIENCE

The work presented in this paper suggests approaches toward maintaining bi-connectivity in data distribution topology in the application layer multicast networks.

The basic principle behind these approaches is that if many nodes form a ring then between any node pair inside the ring, two node-disjoint paths exist. Further if two such rings (or two bi-connected components) share a common link or more than one connected links then those bi-connected components are again bi-connected.

### 3.1 Construction and maintenance of bi-connected mesh:

In the First approach, an algorithm is suggested to construct a bi-connected Mesh over which a distribution tree can be formed.

*Adding newly arriving Node in the Network*
Each new node connects to two already existing nodes in the network. In the beginning, when first node comes, there is no network and new node is the first node in the network (figure 1 a). The second node connects to the first node. The third node connects to the first and second node forming a triangle (figure 1 c). Fourth node connects to any two of the three nodes. The fifth arriving node sees nodes 1 and 4 as having least degree hence it connects to first and fourth node. Mesh formation sequence for first five nodes is shown in figure 1.
The rule followed is that each arriving node connects to two already existing nodes with least degrees. In case two or more than two nodes have the same degree, any two can be used for the connection with equal probability.

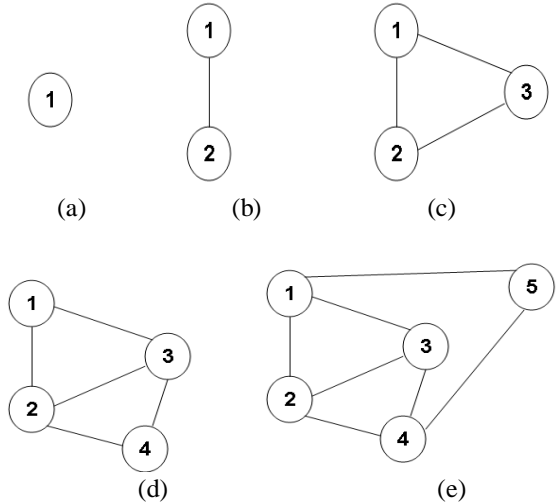

Fig. 1 (a) to (e) Mesh formation sequence as nodes join to the network gradually

Continuing this way, a bi-connected mesh can be formed with any number of nodes. An example mesh with 12 nodes is shown in figure 2.

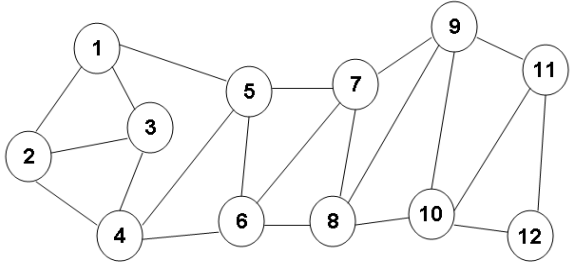

Fig. 2 A bi-connected mesh with 12 nodes

*Analysis of the Mesh so formed*

In the mesh so formed, the average degree of nodes in the mesh, as defined below increases slightly with the number of nodes and reaches to a constant value of 4 as the number of nodes increases to more than 100 (shown in figure 3). The average degree of nodes in a network with N number of nodes can be obtained as below.

$$Average\, degree\, of\, nodes = \frac{\sum_{n=1}^{N} degree\, of\, node\, n}{N}$$

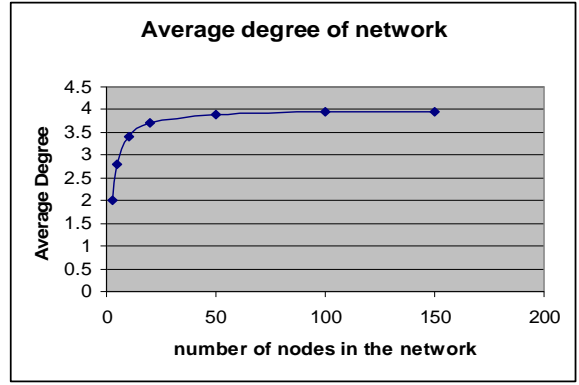

Fig. 3 Average node degree verses network size

Average number of hops travelled for any source destination node pair in the mesh is calculated and plotted in figure 4. It is observed that the number of hops increase linearly with the number of nodes.

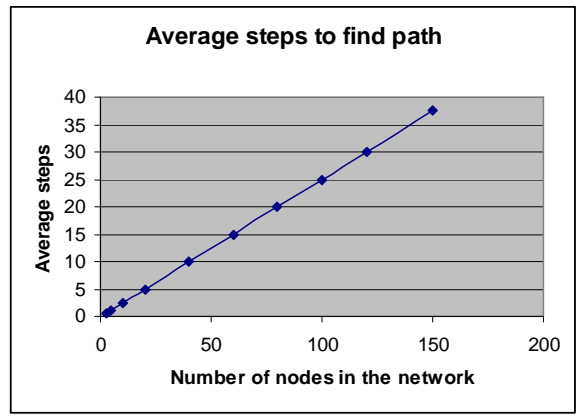

Fig. 4 Average number of hops, averaged over all source destination pairs, considering node 1 as source

*Deleting redundant Links from the Network*

For each existing link in the network, the decision is taken periodically whether to maintain that link or to delete it. If the nodes connected by a link have two node disjoint paths even when the link is removed, that link is declared to be redundant and can be deleted from the mesh if no traffic is passing through it.

*Steps taken in case of node failure*

In case any node fails in the network, the attempt is made to create a ring among the nodes connected to the failed node. Here we assume that every node keeps all its neighbors informed about the other nodes it is connected with. For example in figure 5, all the neighbors of node 7 are aware of its other neighbors. This information is utilized for repairing failures. In case of failure of a node, all the neighbors of the failed node know in advance what are other nodes, to which that failed node was connected. All these neighbors arrange the IDs of neighbors in ascending order. The nodes with highest and lowest ID get connected and the ring is again created among the remaining nodes. For example, in figure 5, if node 7 fails (figure 6), links 5-7, 6-7, 8-7 and 9-7 are removed and a new link 5-9 is created so that all affected nodes form a ring and thus maintaining bi-connectivity.

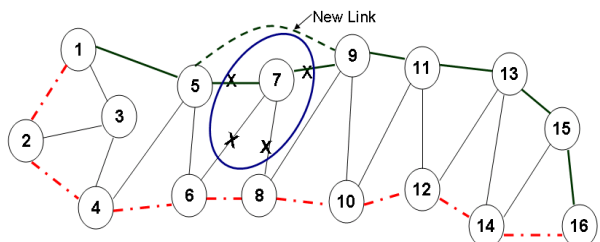

Fig. 5 Addition of new link as node 7 fails

*Advantages and limitations to the approach*

In the mesh so formed, maximum degree that a node can have is limited to 4. But the maximum number of hops between any pair of nodes increases linearly as the number of nodes increase.

**3.2 Approaches for biconnectivity in the tree overlays**

It is also possible to create a tree topology overlay without creating an underlying bi-connected mesh overlay network. For such overlay topologies, we can introduce bi-connectedness property for achieving reliability. In the achieved bi-connected topology, a distribution tree need to be maintained using the algorithm presented in section IV.

*Connected leaf nodes approach*

Biconnectivity in this approach is achieved by connecting all the leaf nodes in pair. By connecting leaf nodes this way, any node pair in the tree gets bi-connectivity, as rings are formed through the leaf nodes. Here we have taken an example of binary tree (shown in figure 6 (a) having 4-level hierarchy. Once the leaf nodes are connected, shown in figure 6 (b), rings are formed between any pair of nodes.

The number of additional links required to achieve bi-connectivity in this approach is (N-1)/2; where N is the number of nodes in the network.

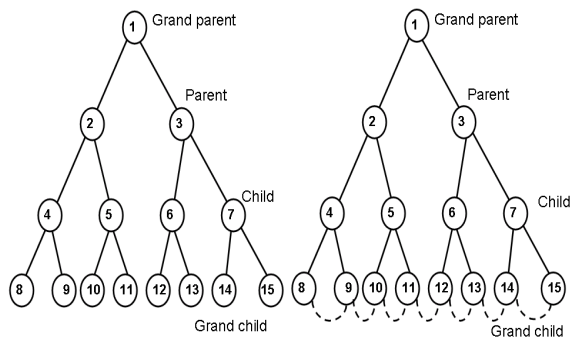

Fig. 6 (a) A binary tree and (b) all leaf nodes are paired to achieve bi-connectivity

*Child-Grandparent approach*

Biconnectivity in this approach is achieved by connecting all the nodes to their grand parent, which provides an alternate path to get the feed in case their parents fail. Wherever it is not possible to connect a node to its grandparent, that node is connected to its siblings. Again an example of binary tree, shown in figure 6 (a), having 4-level hierarchy is considered. Figure 7 (a) shows the original tree along with the new added links, while figure 7 (b) shows only the new added links.

In the child-grandparent approach the number of additional links required is (N-2); where N is number of nodes in the network.

The plot in figure 8 shows that the child-grandparent approach uses almost double the links to achieve Biconnectivity as compared to leaf node approach.

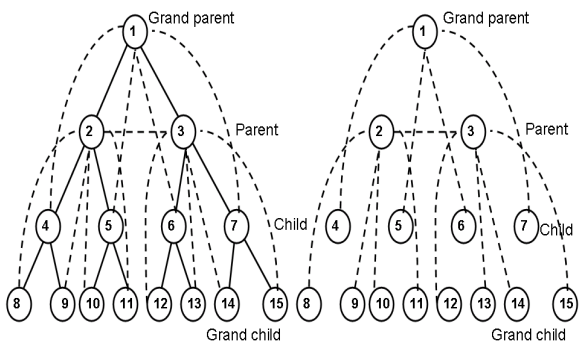

Fig.7 (a) and (b) All nodes are connected to their grandparent and if grandparent is absent then connected to their siblings.

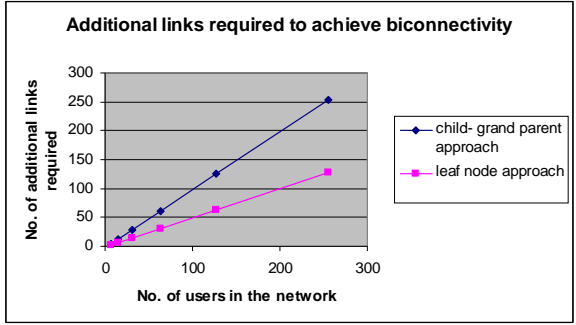

Fig. 8 Number of additional links required to achieve Biconnectivity by different approaches

## IV ALGORITHM FOR DATA DISTRIBUTION TREE OVERLAY IN BI-CONNECTED TOPOLOGY

We assume initially that there exists a mesh containing all the nodes who are participants in a session. The existence of a bi-connected topology is assumed. Further in the case of node or link failure, we assume that repair mechanisms do exist which will restore the bi-connectedness in the topology. This will ensure resilience to any further failures.

**Data Forwarding:**

For distributing the data packets belonging to a live stream, each node maintains a **forwarding table.** Packets coming from the shortest path to source are forwarded to the nodes listed in the table. When a join request is received from a neighbor, its address is added to the forwarding table, and packets are forwarded to it also thereafter.

**Topology Maintenance:**

1. The source of the stream periodically broadcasts signaling packets (beacons) containing source id and sequence number to all the neighbors. The format of signaling packet is shown in Table-1. The sequence number is incremented by the source for every new packet which is broadcasted.

2. Whenever a node gets a packet with new sequence number, its detail { source_id, Sequence_number, from_neighbor (next_hop_to_source)} is included in a table and listed as first best path towards the source (refer Table 2). This packet is then further broadcasted to all the neighbors except to the one from which it was received.

3. When a packet with the same sequence number (the one already listed with the node) arrives at a node from a different neighbor, it is also listed as second best path toward the source in the 'next hop to source table'. Thus the alternate paths to source are always available at every node. This packet is forwarded only to the next hop towards source via the first best path.

4. Any further packets with the same sequence number from any more of the neighbors are simply discarded.

Each node thus maintains the two neighboring nodes which provide the two best paths towards the source. A join request will be forwarded to the best available neighbor node whenever a node wants to receive a data stream.

| From Source ID | Sequence Number | Hop distance from source | Life Time |
|---|---|---|---|
| | | | |

Table 1: Format of signaling packet

**Life Time** is a number that starts with a fixed value proportional to the expected expiry time desired and it decreases with time. It determines the validity of a beacon packet.

The entry ***next hop toward source ID*** helps tracing back the route toward the source. At each node, a record of received sequence numbers is kept in the next hop to source table. This information is stored at each node, so that when a packet (same or different) comes from the same source, it can be compared with the already received packets. Any node receiving the above signaling packet will update these entries before broadcasting it further to all the outgoing links except the one from which it is received. If new and larger sequence number is received from a different neighbor, it indicates that a new and shorter route exists and the next hop entry is updated. The two neighbors from whom first and second packets of same sequence number are received are updated as first next hop and second next hop (to be used in case of failure of reception from first one).

| Source ID | Last Seq. Number received | First Path | | | Second Path | | |
|---|---|---|---|---|---|---|---|
| | | Next hop toward source | Hop distance from source | Timer value | Next hop toward source | Hop distance from source | Timer value |
| | | | | | | | |

Table 2: Format of 'next hop to source' table at each node

Also the life time value from the packet is copied to timer value in the table. The timer value will keep on decreasing with time. With the timeout, the information for that path is deleted. The alternate path information is again entered when beacon packet from same source is received from another neighbor.

Usually 16-bit binary *sequence number* is used and this length can generate 65536 different numbers before repetition occurs. Since the rule is that the entries are updated only if sequence number of received packet is more than the sequence number in the table entry, the timeout allows the update of tables when sequence number 0 is used after completion of a cycle or due to resetting of counter at the source.

**When failure occurs:** Since packets are broadcasted from the source, a downward node can obtain the same packet from different routes earlier or later. This way the node can find many paths to the source, but it stores only two best paths. When a node leaves the overlay, the overlay path to some nodes may fail. Though the transmission can immediately be resumed from the alternate path that is already stored in the next hop to source table.

**New node joining:** When a new node joins the overlay, it will also receive beacon after some time and will know the two best next hop neighbors towards source. It will send a JOIN request packet to best next hop neighbor towards source. If the neighbor is receiving the feed, it will update the forwarding table for multicast and start sending the feed to the new node. If it is not receiving the feed, it will make the entry of new node in the child table for feed and further send the JOIN to the next hop to source. This happens till the JOIN reaches a node (in the worst case to the source) which is getting the feed.

**Adaptive beacon broadcast rate:** Beacon packets broadcasted from the source play an important role in establishing the paths toward the source. In the situation of failures, the beacon broadcast rate can be increased with the increase in the failure rate of nodes so that sudden multiple failures do not interrupt the data delivery. The node that detects a failure of a particular node may request directly to the next best hop towards the source along with the failed node ID and a request to increase the beacon broadcast rate. Then source doubles the rate of broadcast after that particular failure for a certain period. This enables the replenishment of second best path entry in the nodes which are left with only one entry due to the feed failures.

**IV EVALUATION**

Our approach for bi-connectivity maintains an underlying resilient bi-connected mesh. On top of it, a data distribution overlay algorithm is run. Every node maintains the knowledge of two best paths to source. In the situation of failure of one path, feed is made available from the second best path while the broken path is repaired. If no packets are received for certain duration, a node can assume the failure of first feed path, and can send the data delivery request to the second best path. Since there is continuous broadcast of beacon packets, a new path to source will be registered shortly after failure as the second best path. Once the repair of broken path is done, after a while a beacon packet is sure to come from this repaired path and again the two best paths are updated and feed is taken from the first best path.

The time from the instant, a node fails to the instant when a repaired path is registered can be considered as restoration time ($T_{res}$) of the algorithm. The restoration time thus includes the time to detect a failure ($T_{det}$), time to establish additional links at mesh level ($T_{add}$) and time to sense and register optimum path ($T_{reg}$) after repair is done.
$$T_{res} = T_{det} + T_{add} + T_{reg}$$

If the restoration time does not exceed the time between two successive failures, uninterrupted data delivery is maintained from the duplicate paths already stored.

In our approach, the restoration time depends on network diameter, since a new path can be registered only on arrival of fresh beacon packet from the source. This puts the limitation in large networks under high failure rate.

**V CONCLUSION AND FUTURE WORK**

Bi-connectivity of a topology is essential for provisioning of reliable multicast transmission. Three different approaches toward maintaining bi-connectivity have been presented in this paper.

As an extension of the approaches discussed in this paper, for the networks where all nodes are not equally equipped and some nodes are more vulnerable than others, we can opt for tri-connected mesh instead of bi-connected. If the data distribution overlay algorithm (discussed in section III) with the similar extension is applied over this tri-connected mesh, it results in three best paths to source. Since only signaling information is required to be propagated to maintain these paths, only a small bandwidth is consumed and hence it proves as an viable option for vulnerable nodes to maintain three best paths.

As a further extension, a large (heterogeneous) network can be divided into different protective zones; where different zones can be protected with bi-connected, tri-connected or n-connected mesh, and 2,

3 or n best paths can be maintained depending on the vulnerability of different zones.